\documentclass{WileyMSP-template}
\usepackage{pdfpages}
\begin{document}
\pagestyle{fancy}

\title{In-line Dispersion Control of Femtosecond Laser Pulses by Multilayer Huygens' Metasurfaces}

\maketitle


\author{Anna Fitriana,*}
\author{Katsuya Tanaka,}
\author{Roland Schiek,}
\author{Thomas Pertsch,}
\author{Dragomir Neshev,}
\author{Isabelle Staude}




\begin{affiliations}

A. Fitriana, K. Tanaka, I. Staude\\
Institute of Solid State Physics\\
Friedrich-Schiller-University Jena\\
Max-Wien-Platz 1, 07743 Jena, Germany\\
Email: anna.fitriana@uni-jena.de\\
\vspace{5mm} 
A.Fitriana, K. Tanaka, T. Pertsch, I. Staude\\
Institute of Applied Physics Physics\\
Friedrich-Schiller-University Jena\\
Albert-Einstein-Straße 6, 07745 Jena, Germany\\
\vspace{5mm} 
A. Fitriana, R. Schiek, D. Neshev\\
ARC Centre of Excellence for Transformative Meta-Optical Systems (TMOS)\\
Department of Electronic Materials Engineering, Research School of Physics\\
Australian National University\\
Canberra, 2600 ACT, Australia\\
\vspace{5mm} 
T. Pertsch\\
Fraunhofer Institute for Applied Optics and Precision Engineering IOF\\
Albert-Einstein-Straße 7, 07745 Jena, Germany
\end{affiliations}

\keywords{dispersion engineering, Huygens' regime, multilayer metasurfaces, pulse compression, ultrafast photonics}

\begin{abstract}
Dispersion is a key limiting factor in ultrafast systems as it modifies the temporal structure of ultrashort optical pulses. Conventional dispersion compensation techniques, though widely used, rely on a combination of several bulky optical components and complex, long propagation paths, posing challenges for integration into compact photonic platforms. All-dielectric resonant metasurfaces operating in the Huygens' regime of spectrally overlapping electric and magnetic dipolar resonances offer a promising route to low-loss dispersion control at the sub-wavelength scale; however, the achievable dispersion from a single layer remains fundamentally limited by its phase gradient and spectral bandwidth. Here, we experimentally demonstrate in-line dispersion control of femtosecond laser pulses using a multilayer architecture of dielectric Huygens’ metasurfaces, without requiring spatial separation of their spectral components, as typically employed in Fourier-plane dispersion compensation schemes. The phase dispersion accumulates across stacked layers approaching a total of $\sim$8$\pi$ corresponding to a maximum group delay dispersion of around 3969 fs\textsuperscript{2}. This enables efficient compensation of the input chirp and results in a significant compression of femtosecond laser pulses in the telecom wavelength range. Importantly, the metasurface stack functions as a single compact transmissive element, which is simply inserted into the beampath. Overall, our results establish a compact, scalable platform for integrated pulse compression and ultrafast dispersion engineering.
\end{abstract}

\section{Introduction}\label{intro}
Ultrashort laser pulses enable direct probing and control of ultrafast phenomena, underpinning advances across ultrafast spectroscopy, time-resolved imaging, optical communication, and nonlinear optics \cite{brabec2008strong,zewail2000femtochemistry,dudovich2002single,drexler2004ultrahigh}. Realizing the full potential of such pulses, however, requires maintaining their Fourier transform-limited durations, as this maximizes their peak intensity and thereby a range of nonlinear light-matter interactions that are essential for various applications \cite{sidick1995ultrashort,lozovoy2005systematic}. Achieving transform-limited pulse durations critically depends on careful control of dispersion, particularly group delay dispersion (GDD) in optical systems, since even moderate GDD can significantly increase pulse duration and distort pulse envelope \cite{simon1989ultrashort,wise2008high}. Compensating dispersion is therefore one of the central challenges in ultrafast photonics \cite{walmsley2001role,dar2017chromatic}.

In realistic optical systems, dispersion accumulates rapidly through transmissive optics, substrates, and amplification stages, often reaching tens to thousands of fs\textsuperscript{2} \cite{strickland1985compression,brabec2000intense}. Conventional dispersion compensation, such as angularly dispersive elements \cite{bor1985group,treacy1969optical}, reflective compressors \cite{dombi2005pulse,chia2025multi}, fiber-based platforms \cite{poole1994optical,gruner2005dispersion}, as well as Fourier plane pulse shapers based on spatially-modulated light \cite{weiner2000femtosecond, lee2006optical} or acousto-optic modulators \cite{xie2005dispersion,catanese2021acousto}, are widely used to restore ultrashort durations. While highly effective, many of these approaches typically rely on several bulky components, long optical paths between them, and sensitive angular alignment. Such requirements inevitably increase system complexity, maintenance requirement, size and weight. Additionally, they introduce beam deviations, therefore hindering their integration in compact photonic platforms.

More recently, planar metasurfaces have emerged as a promising alternative for dispersion engineering in ultrathin optical elements. By tailoring subwavelength structures, metasurfaces can impose engineered spectral phase responses through mechanisms such as geometric phase modulation \cite{gokce2014femtosecond} and resonance-induced dispersion \cite{rahimi2015femtosecond,ossiander2021slow,geromel2023compact}. Furthermore, when metasurfaces are implemented in a Fourier plane configuration, the spectral phase manipulation can be spatially and individually encoded in metasurface pixels, allowing programmable amplitude and phase modulation, as well as wide-range spectral control \cite{divitt2019ultrafast, chen2022synthesizing, chen2023shaping, tanaka2025femtosecond}. Although these approaches offer considerable flexibility for ultrafast pulse shaping, they retain long optical paths inherent to Fourier-optical systems that pose challenges for fully integrated devices. 

A complementary strategy to Fourier plane configuration is to directly use the GDD inherent to a resonant metasurface for shaping ultrashort pulses. Specifically, all-dielectric Huygens' metasurfaces \cite{decker2015high} utilize spectrally overlapping electric and magnetic dipoles to achieve nearly-unity transmission with a full $2\pi$ phase coverage. The strong spectral variation of the transmittance phase around the resonance frequency naturally produces significant GDD, arising from the steep curvature of the wavelength-dependent phase \cite{chen2023dispersion}. However, achieving large GDD using a single resonant element generally requires resonances with high quality factor (Q-factor) \cite{barati2022optical}, which simultaneously narrow their operational bandwidth and fundamentally impose a trade-off between phase-gradient and bandwidth. 

\begin{figure*}
\centerline{\includegraphics[width=372pt,height=19pc]{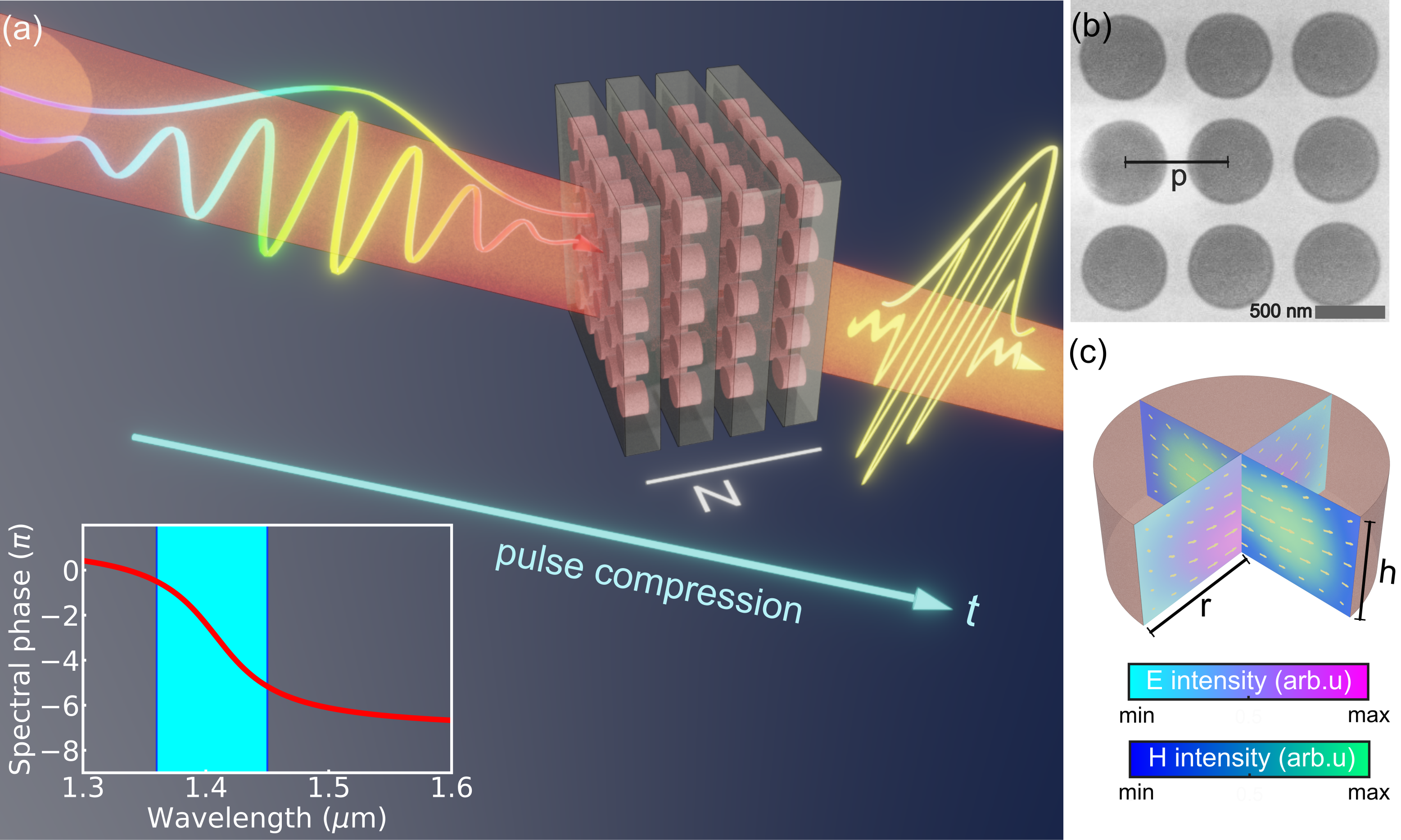}}
\caption{Dispersion compensation of a femtosecond laser pulse using a multilayer stack of all-dielectric Huygens' metasurfaces. (a) A chirped input pulse is incident onto the silicon nanodisk metasurfaces, resulting in temporal pulse compression. The transmission spectral phase accumulates during the propagation through several metasurface layers \textit{N}. The cyan shaded region indicates the wavelength range of the input pulse, where the pulse distortion occurs. (b) Scanning electron micrograph (SEM) of the fabricated silicon metasurface. (c) Simulated near-field intensity distributions, \textbar\textit{E}/\textit{E}\textsubscript{0}\textbar\textsuperscript{2} and \textbar\textit{H}/\textit{H}\textsubscript{0}\textbar\textsuperscript{2} at a wavelength of 1.40 {\textmu}m, highlighting the electric and magnetic Mie-type dipolar resonances, respectively. Arrows indicate the tangential field vectors of \textit{E} and \textit{H} in the corresponding planes. The geometrical metasurface parameters used in this simulation are \textit{r} = 251 nm, \textit{h} = 225 nm, and \textit{p} = 754 nm, corresponding to the effective geometries confirmed both by the complex transmission simulation and SEM inspection (see SI Section S1).}
\label{fig:concept}
\end{figure*}

Here, we overcome this challenge through in-line phase-dispersion control by a multilayer arrangement of identical dielectric Huygens' metasurfaces (see Figure~\ref{fig:concept}a). The transmitted phase accumulates additively across the successive layers, which leads to substantial GDD without relying on high-Q resonances. Experimentally, we obtained an aggregate phase coverage of $\sim$8$\pi$ using a four-layer stack, corresponding to a maximum GDD magnitude of 3969 fs\textsuperscript{2}. The resulting structure functions as a single transmissive optical element of compact thickness that can be simply inserted into the target beampath. As such, our approach provides a direct control over the pulse shape and establishes a scalable route toward ultrathin dispersion engineering. 

\section{Results}\label{results}

An artistic impression of dispersion compensation based on a stacked arrangement of Huygens' metasurfaces is shown in Figure~\ref{fig:concept}. Each metasurface layer consists of silicon (Si) nanodisks with radius \textit{r} and height \textit{h}, arranged in periodic square arrays with period \textit{p} and embedded within a glass matrix. Numerical simulation of the optical near-fields for excitation with a normally incident plane wave at a wavelength of 1.40 {\textmu}m shows strong field confinement corresponding to two dominant resonant modes of electric and magnetic dipolar character, respectively. The spatial coexistence and spectral overlap of these modes give rise to a rapid phase variation due to the interfering resonances. Specifically, under Huygens' condition where the overlapping electric and magnetic dipole responses are balanced in amplitude and spectral bandwidth, a destructive interference occurs in the backward direction, leading to suppressed reflection and ideally near-unity transmission. Details on the numerical simulation and fabrication process of the Huygens' metasurfaces can be found in the Supporting Information (SI) Sections S1 and S2. 

The influence of a highly transmissive optical system on the time-domain profile of ultrashort laser pulses is captured by the frequency-dependent spectral phase $\varphi(\omega)$. The first derivative of the spectral phase with respect to the angular frequency defines the group delay, GD $= \frac{ \mathrm{d}\varphi(\omega)}{\mathrm{d}\omega}$, which describes the arrival time of the pulse envelope. The second derivative, known as group delay dispersion, GDD $= \frac{\mathrm{d}^2 \varphi(\omega)}{\mathrm{d}\omega^2}$, quantifies the frequency-varying group delay, where different spectral components propagate at different velocities and, as a result, leads to distortion or broadening of the pulse envelope. To compensate for this broadening, the GDD introduced by our metasurface is designed to counteract the GDD of the incident pulse. 

The interaction between the incident pulse and the metasurface is described in the frequency domain by a complex transfer function $H(\omega)$, such that the output spectrum satisfies $\mathcal{F}\{{f_{\mathrm{out}}(t)}\}$, given by 
\begin{equation}
\mathcal{F}\{{f_{\mathrm{out}}(t)}\} = H(\omega)\mathcal{F}\{{f_{\mathrm{in}}(t)}\}
\label{eq: pulse_shape}
\end{equation}
where $\mathcal{F}$ denotes the Fourier transform operation and $f_{\mathrm{in}}(t)$ and $f_{\mathrm{out}}(t)$ are the complex electric fields in the temporal domain of the input and transmitted pulses, respectively. The transfer function $H(\omega)$ encodes both amplitude and phase of the transmission function of the metasurface. 

\begin{figure*}
\centerline{\includegraphics[width=372pt,height=24pc]{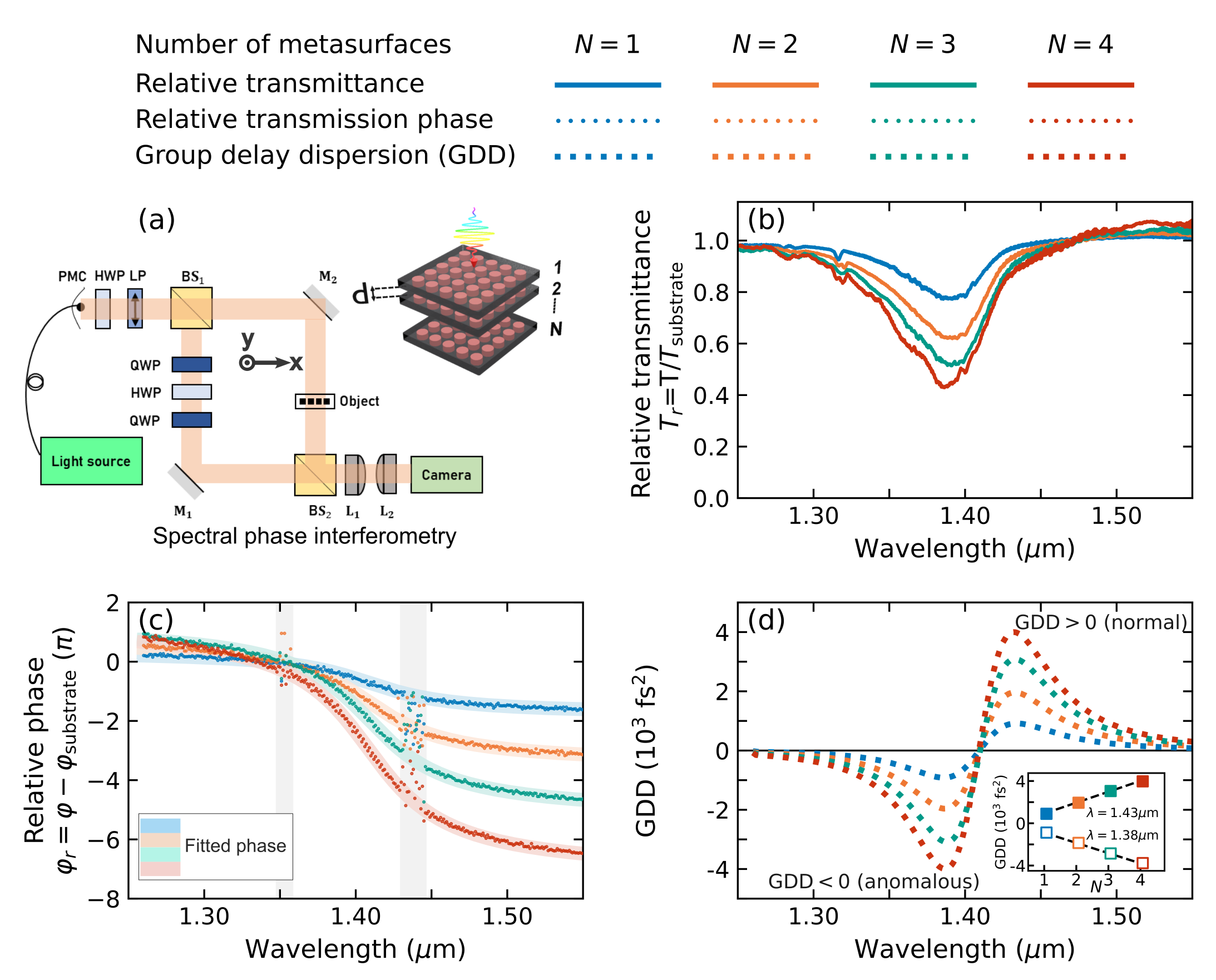}}
  \caption{Dispersion characteristics of stacked silicon Huygens' metasurfaces. (a) \textbf{Left}: schematic of the spectral phase interferometry setup used to determine the complex transmission function of the fabricated Huygens' metasurfaces. \textbf{Right}: a multilayer configuration of metasurfaces consists of \textit{N} identical layers separated by a distance \textit{d}. (b) Experimentally measured relative transmittance of \textit{N}-layer stacks. (c) Retrieved spectral phase for different \textit{N}, where the experimental results (dots) are fitted by a modified arctan function (thick solid lines) derived from a Lorentzian resonance for each \textit{N}. The gray-shaded areas indicate the frequency ranges in which our laser source could not operate stably, resulting in large measurement fluctuations. (d) Calculated GDD spectra derived from the fitted phase in (c) are depicted by block-dashed lines, exhibiting positive and negative dispersion regimes. The GDD values as a function of \textit{N} at wavelengths of 1.38 and 1.43 {\textmu}m are presented as open and filled squares in the inset, respectively. Linear fits (dashed dark) yield: GDD = -0.97\textit{N} + 0.08 at 1.38 {\textmu}m and GDD = 1.03\textit{N} + 0.11 at 1.43 {\textmu}m, both exhibiting fitting variance \textit{R}\textsuperscript{2} \textgreater 0.99.}
  \label{fig:GDD}
  \end{figure*}

\begin{figure*}[!t]
\centerline{\includegraphics[width=372pt,height=25pc]{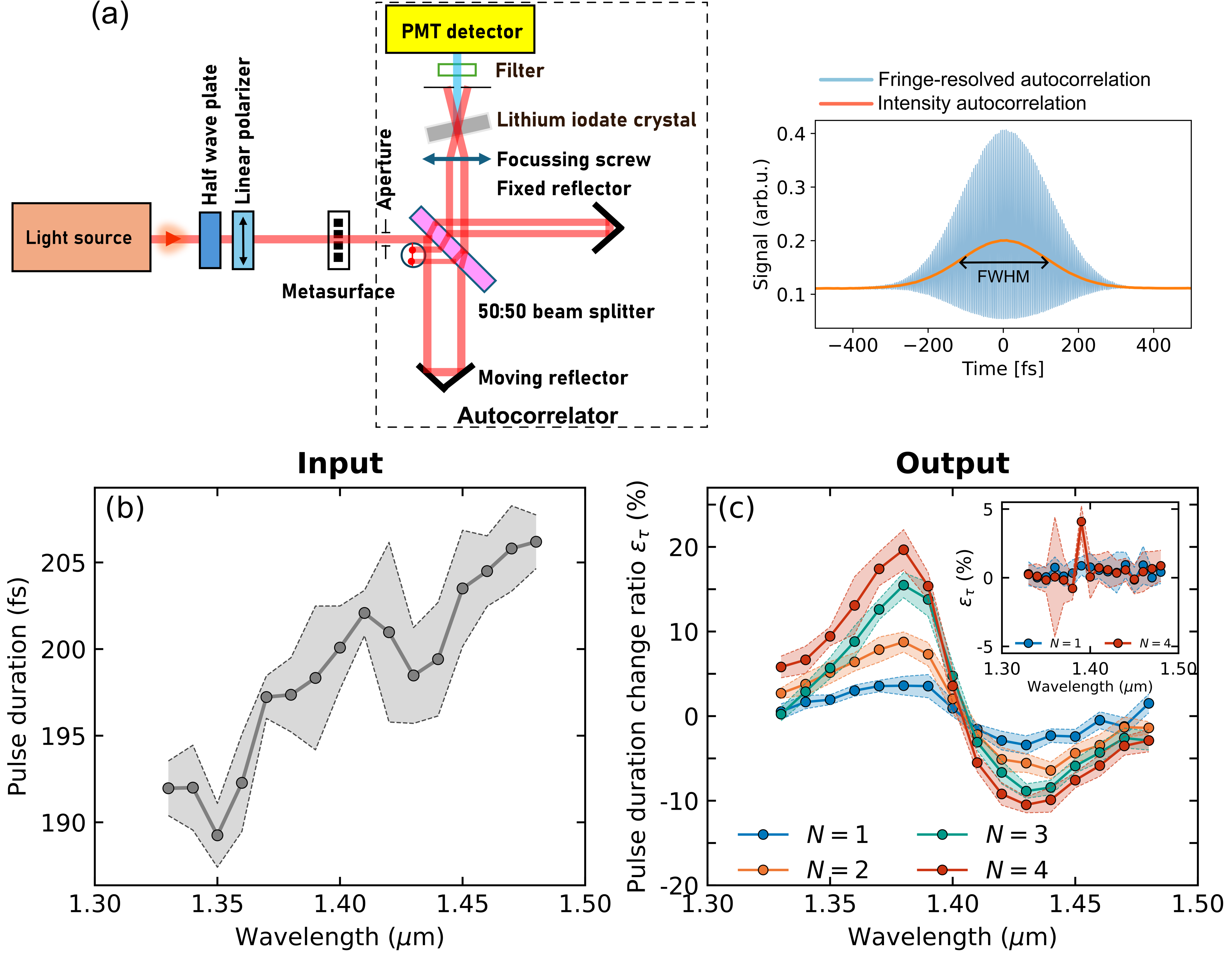}}
  \caption{Intensity autocorrelation measurement of femtosecond pulses. (a) \textbf{Left}: schematic of the pulse characterization setup using an autocorrelator (APE, pulseCheck NX). \textbf{Right}: fringe-resolved (interferometric) autocorrelation trace and the corresponding intensity autocorrelation. The latter is obtained by either experimentally inserting an NIR filter or mathematically removing fast oscillating components in the Fourier domain. (b) Measured temporal duration of the input pulses. The shaded region indicates the measurement uncertainty due to pulse energy fluctuation in the laser output. (c) Relative change in the temporal duration for pulses transmitted through the four different metasurface stacks with respect to the duration of input, calculated based on Equation (\ref{eq:var_tau}). Positive and negative values of $\varepsilon_\tau$ indicate temporal broadening and compression, respectively. Similar measurements were performed for the stacked case of 1-mm-glass substrates as illustrated in the inset, showing negligible temporal modification of the transmitted femtosecond pulses.}
  \label{fig:autocorr}
  \end{figure*}

The dispersion characteristics of the Si Huygens' metasurfaces were systematically investigated for different numbers \textit{N} of stacked metasurfaces, as illustrated in Figure~\ref{fig:GDD}a. Adjacent layers were arranged using a customized sample holder with a fixed separating distance \textit{d}=1.3$\pm$0.1 mm. This interlayer spacing is several orders of magnitude larger than the coherence length of the laser source used in the transmission measurements ($L_c\sim$60 {\textmu}m), which effectively suppresses the contribution of near-field coupling and coherent interference between neighboring metasurfaces. 
 
Figure~\ref{fig:GDD}b presents the transmittance $T_r$ of the Huygens metasurface stacks for $N=1,2,3,$ and $4$ defined relative to the corresponding $N$-stack glass substrates, written as $T_\mathrm{r} = T/T_{\mathrm{substrate}}$. Here, $T$ is the absolute transmittance and $T_\mathrm{substrate}$ is the transmittance of the glass-substrate stack. Notable transmittance minima appear near 1.39 {\textmu}m wavelength, originating from the spectrally overlapping electric and magnetic modes excited in the silicon nanodisks. Spectral overlap of electric and magnetic dipolar resonances is commonly exploited to achieve impedance-matching and thus near unity transmission in a Huygens' metasurface. In practice, however, this condition is sensitive to the geometric variations of the nanodisks caused by fabrication tolerances, and therefore introduces small residual reflections around the resonance. The resonance exhibits a full-width at half maximum (FWHM) of $\sim$60 nm, indicating a relatively broad resonance linewidth of the overlapping dipole modes. We observe that as the number of layers \textit{N} increases, the relative transmittance $T_r$ at the resonant dip near 1.39 {\textmu}m decreases from 0.78 for \textit{N} = 1 to 0.45 for \textit{N} = 4. These values correspond to the absolute transmittance $T$ of 0.72 and 0.34 for \textit{N} = 1 and \textit{N} = 4, respectively. Details of the transmission characterization results are provided in SI Figure S2. Notably, the 4-stack glass substrate alone exhibits a transmittance of 0.85, indicating that the reflections at the glass interfaces introduce substantial transmission losses, which further contribute to the difference between $T_r$ and $T$. The transmission values could potentially be increased by improving fabrication precision and suppressing interface reflections, for example, by applying anti reflection coatings to back surfaces of the glass substrates or immersing the stack in an index-matched medium.

The spectral phase of the relative transmission $\varphi_r$ was then retrieved experimentally and plotted in Figure~\ref{fig:GDD}c (dots). To obtain smooth and physically meaningful GDD spectra of the metasurfaces, we fitted the phase response using a Lorentzian resonance profile. Within this framework, the phase can be written as $\varphi_r (\omega) = a+b$ arctan$\bigg[\frac{\omega-\omega_0}{\gamma/2}\bigg]$, where  $\omega_0$ is the resonant frequency and $\gamma$ is the effective decay rate of the resonance. The justification for using this specific fitting function is detailed in SI Section S3. As shown in Figure~\ref{fig:GDD}c, the fitted curve captures the dispersive behavior of the resonances with high fidelity. 

The extracted fitting parameters evaluated for each layer configuration are summarized in Table \ref{tab:phase}. Here, $a$ corresponds to the phase value at the resonance dip, located around $\lambda_0=1.41$ {\textmu}m or $\omega_0=1.34 \times$10\textsuperscript{15} s\textsuperscript{-1}, while $b$ quantifies the phase coverage across the spectral range of interest. The resonance linewidth is characterized by the effective decay rate $\gamma$. Notably, $\gamma$ values are nearly invariant with increasing \textit{N}, indicating that stacking does not alter the spectral bandwidth of the resonance. For a four-layer configuration, \textit{N} = 4, a total phase accumulation of around $8\pi$ is experimentally achieved. 

\begin{table}[h]
\caption{Coefficients of transmission phase function in \textit{N}-layer metasurfaces}
\centering
\begin{tabular}{c c c c c}
\hline
\textit{N}& $a$ & $b (\pi)$ & $\gamma$ (10\textsuperscript{13} s\textsuperscript{-1})& $\omega_0$ (10\textsuperscript{15} s\textsuperscript{-1}) \\
\hline
1 & -0.72 & 2.20 & 7.89 & 1.34 \\
2 & -1.38 & 4.33 & 7.56 & 1.34 \\
3 & -2.03 & 6.49 & 7.39 & 1.34 \\
4 & -3.00 & 8.60 & 7.47 & 1.34 \\
\hline
\end{tabular}
\label{tab:phase}
\end{table}

Further, we calculated the intrinsic GDD of the metasurface stacks shown in Figure~\ref{fig:GDD}d. The dispersive behavior of a Lorentzian resonance is characterized by a rapid spectral phase transition accompanied by a change in the curvature, which corresponds to a sign flip of the GDD. Negative GDD values are observed in the 1.30--1.40 {\textmu}m spectral range, at which the lower-frequency (red) components of pulses experience larger delay. Positive GDDs appear between 1.41--1.55 {\textmu}m where the higher-frequency (blue) components are delayed more strongly. At 1.38 {\textmu}m wavelength, the single metasurface yields a GDD of -881 fs\textsuperscript{2}, which increases in magnitude to -3770 fs\textsuperscript{2} in a four-layer stack (see filled squares in the inset). In contrast, at 1.43 {\textmu}m, the retrieved GDD values (open squares) are +899 fs\textsuperscript{2} and +3969 fs\textsuperscript{2} for \textit{N} = 1 and 4, respectively. The GDD magnitude increases approximately linearly with the number of metasurface layers, as confirmed by the linear fits shown in the inset of Figure~\ref{fig:GDD}d. This scalable behavior demonstrates that the dispersion imparted by a stack of Huygens' metasurfaces can be systematically engineered through layer-by-layer assembly, thus highlighting their potential as compact, tunable dispersion-compensating elements.

Next, we investigated the temporal dynamics of femtosecond pulses interacting with Huygens' metasurfaces. Autocorrelation measurements were performed using negatively chirped pulses with central wavelengths ranging from 1.33 to 1.48 {\textmu}m. The pulses were generated using an optical parametric oscillator (OPO, Spectra-Physics Inspire HF100) pumped by a mode-locked titanium:sapphire laser (Mai Tai, repetition rate of 80 MHz). The temporal durations were first estimated by fitting the measured intensity autocorrelation to the calculated autocorrelation of a Gaussian pulse envelope. This choice was motivated by the near-Gaussian profile of the pulses and ensured a minimal root-mean-square (RMS) residual error between the measured and modeled autocorrelation (see SI Section S6). Figure~\ref{fig:autocorr}a shows the schematic of the autocorrelation measurement setup. The estimated pulse durations are shown in Figure~\ref{fig:autocorr}b together with the measurement uncertainties indicated by the shaded area. These variations primarily reflect time-dependent intensity fluctuations in the laser output and were statistically estimated from nine consecutive autocorrelation measurements. 

Figure~\ref{fig:autocorr}c presents the relative change of pulse duration, $\varepsilon_\tau$, after propagating through the metasurfaces, which is defined as: 
\begin{equation}
\varepsilon_\tau (\%) = \frac{\tau_{\mathrm{out}}-\tau_{\mathrm{in}}}{\tau_{\mathrm{in}}} \times 100 \%,
\label{eq:var_tau}
\end{equation}
where $\tau_{\mathrm{in}}$ and $\tau_{\mathrm{out}}$ denote the full width at half maximum (FWHM) of the temporal durations for the input and transmitted pulses, respectively. Positive values of $\varepsilon_\tau$ are observed in the wavelength range of 1.33--1.40 {\textmu}m, indicating pulse broadening, whereas the negative values in the 1.42--1.55 {\textmu}m range correspond to pulse compression. For a four-layer stack (\textit{N} = 4), a maximum pulse compression of $\sim$10.5$\pm0.9 \%$ is achieved at 1.43 {\textmu}m, while the largest pulse broadening of $\sim$19.7$\pm2.4 \%$ occurs at 1.38 {\textmu}m. 

Notably, the evolution of $\varepsilon_\tau$ closely follows the spectral profile of the GDD of the metasurface stacks, but with opposite sign. This feature clearly indicates that the dispersion of the metasurfaces governs the temporal pulse shaping. Specifically, the phase curvature imposed by the resonant metasurfaces introduces wavelength-dependent phase dispersion that either compensates for or enhances the initial pulse chirp. Here, we show that the negative dispersion of the input is progressively compensated by the positive GDD of the metasurfaces, leading to pulse compression. Conversely, when the input and metasurface GDD have the same sign, the accumulated dispersion increases and results in pulse broadening.

As shown in the inset of Figure~\ref{fig:autocorr}c, the corresponding autocorrelation measurements for a stack of 1-mm-thick glass substrates exhibit negligible variation in $\varepsilon_\tau$. The only notable exception occurs at a wavelength of 1.39 {\textmu}m, where a stack of four substrates yields $\varepsilon_\tau =$ 4.1$\pm$1.0$\%$. This singular result is likely due to a measurement error caused by slight misalignment in the autocorrelation setup, which is further amplified by multiple reflections between adjacent glass substrates. Importantly, the dispersion of the millimeter-thick glass substrates, is insufficient to produce measurable temporal distortion in 200-fs pulses. 

To further investigate the temporal structure of the femtosecond laser pulses, we performed second-harmonic frequency-resolved optical gating (SH-FROG) measurements. In this case, we performed FROG measurements using positively chirped pulses which were generated from a titanium:sapphire laser (Coherent, Chameleon Ultra II) and the corresponding laser output wavelength was further adjusted using an OPO (APE, Chameleon Compact). To demonstrate temporal pulse compression, we exploited the negative dispersion regime of the resonant metasurface with GDD magnitude reaching its maximum at around 1.38 \textmu m (see Figure~\ref{fig:GDD}d). We thus recorded FROG trace for pulses centered at 1.375 $\pm$ 0.003 \textmu m before and after transmission through a stack of four Huygens’ metasurfaces, in the following denoted as input and output, respectively. The retrieval process of the SH-FROG data enables the complete reconstruction of the complex electric field in both the temporal and spectral domains.

\begin{figure*}
\begin{center}
  \includegraphics[width=352pt,height=37pc]{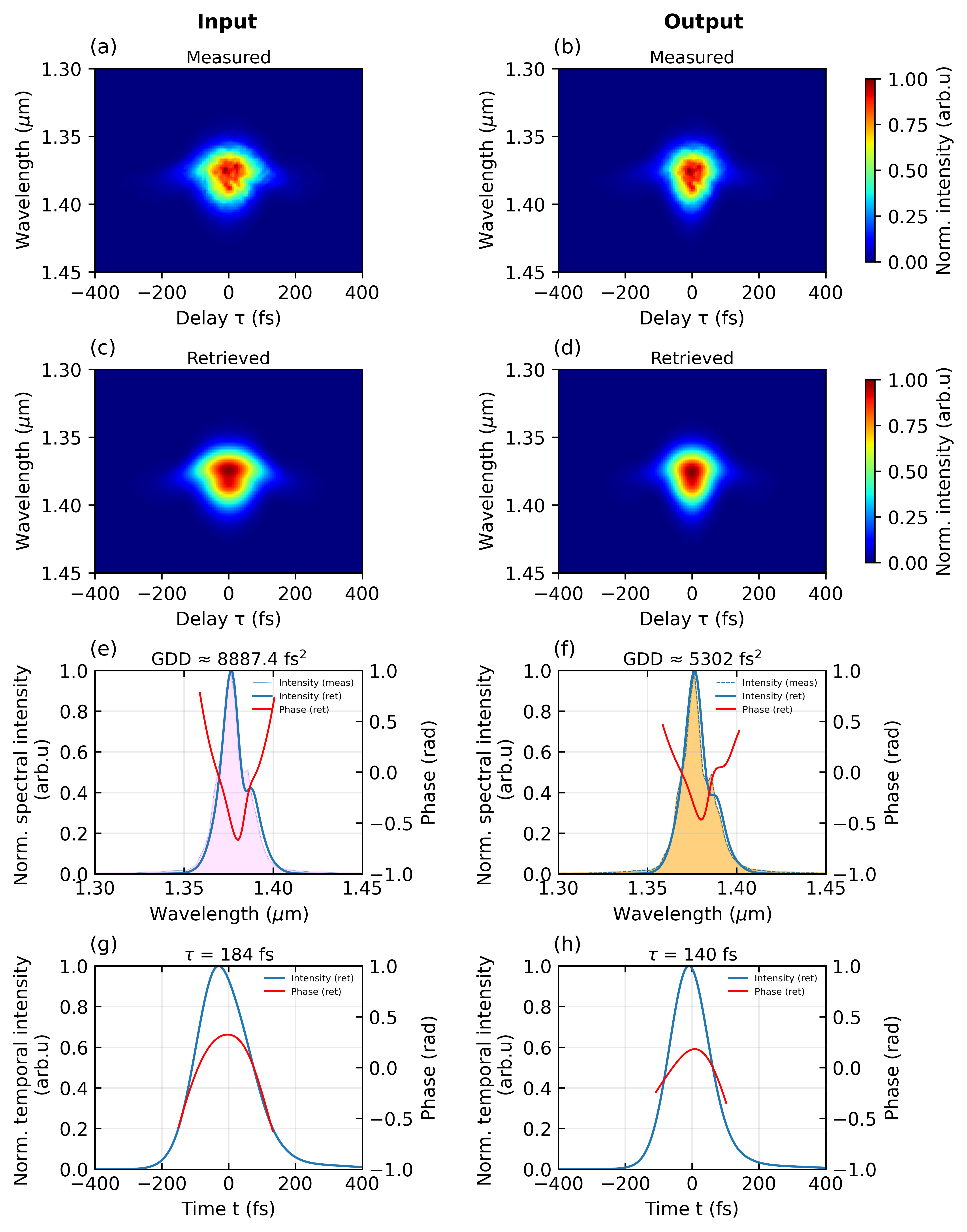}
    \caption{Second harmonic frequency-resolved optical gating (SH-FROG) characterization of the input and output pulses centered at 1.375 $\pm$ 0.003 {\textmu}m. (a,b) Experimentally measured FROG spectrogram of the input and output pulses, respectively. (c,d) Retrieved FROG spectrogram obtained using a phase-retrieval algorithm \cite{delong1994pulse,delong1994improved}, showing a good agreement with the measured traces, with a retrieval error below $1\%$. (e,f) Normalized spectral intensity and phase, retrieved for the input and output pulses. Filled regions indicate the corresponding measured spectra. (g,h) Retrieved temporal intensity and phase, demonstrating pulse compression of the output pulse after propagating through metasurfaces.}
  \label{fig:FROG}
  \end{center}
  \end{figure*}
  
Figures~\ref{fig:FROG}a and b present the FROG traces of the input and output pulses, respectively, normalized to their respective peak intensities. Prior to FROG retrieval, the raw traces were pre-processed by performing a background subtraction and noise removal. Also, a validity of the measured FROG trace was verified by a comparison of the FROG frequency marginal with the autoconvolution of the spectrum independently measured using a spectrometer (SI Section S5). Figures~\ref{fig:FROG}c and d present the retrieved FROG spectrograms for the input and output pulses, respectively, demonstrating a good agreement with the measured source traces in Figure~\ref{fig:FROG} a and b. This is also quantified by a FROG error, $G$, below 1$\%$. Details on the error calculation are provided in SI Section S7. 

The retrieved intensity and phase spectra of the pulses are shown in Figures~\ref{fig:FROG}e and f, represented by the blue and red curves, respectively. The retrieved spectral intensities are in excellent agreement with the independently measured spectra, confirming the reliability of our FROG measurements. To quantify the dispersion response of the resonant metasurfaces, we evaluated the GDD of the input and output pulses at the center wavelengths. The GDD decreased from +8887 fs\textsuperscript{2} for the input pulse to +5302 fs\textsuperscript{2} for the output pulse, corresponding to a reduction of approximately 3585 fs\textsuperscript{2} after interaction with the metasurfaces. This directly evidences compensation of the initial pulse chirp by the metasurface-induced dispersion and agrees well with the independently measured GDD of -3417 fs\textsuperscript{2} reported at 1.375 \textmu m from the spectral phase interferometry (see~Figure \ref{fig:GDD}d). Further, the retrieved temporal intensity and phase profile in Figures~\ref{fig:FROG}g and h reveal substantial pulse compression, with the pulse durations decreasing from 184 fs to 140 fs after transmission through the stack of four metasurfaces. 

\section{Conclusion}\label{conclusion} 
In summary, we experimentally demonstrate dispersion engineering through spectral phase manipulation in multilayer dielectric Huygens' metasurfaces. Operating in the Huygens' regime allows for rapid change in the spectral phase profile with a 2$\pi$ phase variation while preserving high transmission intensity. By stacking nominally identical metasurface layers, the transmitted phase accumulates linearly across successive interfaces, resulting in a proportional enhancement of the group delay dispersion. This scalable phase accumulation provides a simple yet powerful mechanism for tailoring the temporal evolution of femtosecond pulses through control of the number of metasurface layers for both pulse compression and broadening. The multilayer architecture offers a robust and fabrication-friendly route to programmable dispersion without increasing the structural complexity. More broadly, our work extends the functionality of Huygens' metasurfaces beyond spatial wavefront control to temporal wavefront engineering, establishing stacked metasurfaces as a compact and scalable platform for dispersion management and ultrafast pulse shaping.

\section{Experimental Sections}
\label{sect:method} 
\subsection{Spectral phase interferometry}
\label{sect:interferometric} 
The complex transmission function of the metasurface was characterized using Mach-Zehnder phase-shift interferometry (see Figure~\ref{fig:GDD}(a)). The light source from a tunable narrow-linewidth infrared laser (Santec, TSL-210F) was collimated using a reflective collimator (Thorlabs, RC02) and horizontally polarized using a combination of half wave plate and linear polarizer. The beam was then split into reference and sample arms using a 50:50 non-polarizing beam splitter BS\textsubscript{1} (Thorlabs, BSW23R). Phase modulation was implemented in the reference arm using a rotating half wave plate (Thorlabs, SAHWP05M-1700) placed between two quarter wave plates (Thorlabs, SAQWP05M-1700). We applied 16 discrete phase steps between 0 and $2\pi$ in the half wave plate in order to suppress noises and measurement errors and increase retrieval accuracy \cite{malacara2007optical}. The two beams were further interfered in the second beam splitter BS\textsubscript{2}. The resulting interferograms were relayed through a 4f imaging system (L\textsubscript{1}, f = -100 mm and L\textsubscript{2}, f = 200 mm) and recorded onto an InGaAs camera (Xenics, Xeva). Spectrally resolved amplitude and phase were retrieved using a standard phase reconstruction algorithm (see SI Section S4).  

\subsection{Temporal Pulse Characterization}
\label{sect:pulse characterization}  
\subsubsection{Interferometric Autocorrelation Measurement}
\label{sect:autocorrelation}  
The temporal durations of femtosecond pulses were characterized using a commercial autocorrelator (APE, PulseCheck NX100). A schematic of the interferometric autocorrelation measurement setup is shown in Figure~\ref{fig:autocorr}a. In this configuration,  the source pulse was split into two replicas where the relative path difference between them was controlled by a scanning stage. The two replicas were subsequently recombined into a collinear beam and focused into a second harmonic generation crystal, where the nonlinear interaction produced an interferometric autocorrelation signal. The resulting interference fringes were recorded as a function of the relative time delay.

The autocorrelation measurements were carried out for pulses with central wavelength spanning in the 1.33--1.58 {\textmu}m wavelength range. The femtosecond pulses were generated by an optical parametric oscillator (Spectra-Physics, Inspire HF100) pumped by an 800-nm mode-locked titanium-sapphire laser (Mai Tai, repetition rate of 80 MHz). 

The intensity autocorrelation trace was obtained by applying a low-pass filter to the interferometric signal to remove the rapidly oscillating components of the second harmonic fringes. Finally, assuming a Gaussian pulse profile, the temporal pulse durations, $\tau$, were determined from the FWHM of the intensity autocorrelation signal, $\tau_{\mathrm{IA}}$, according to the following standard relation $\tau = 0.707 \tau_{\mathrm{IA}}$ \cite{diels2006ultrashort}. 

\subsubsection{Second Harmonic Frequency-Resolved Optical Gating Measurement}
\label{sect:SH FROG}  
For a complete reconstruction of both temporal and spectral intensity and phase, second harmonic frequency-resolved optical gating (SH FROG) measurements were performed using a compact single-shot FROG device, Grenouille (Swamp Optics, 15-40 USB). Unlike conventional SH FROG systems, this device integrates a cylindrical lens, a Fresnel biprism, a second harmonic crystal, and an imaging lens into a compact architecture, which substantially eliminates the complexity associated with the SH FROG setup \cite{o2001highly}. The recorded FROG traces were retrieved using an iterative retrieval algorithm \cite{delong1994pulse,delong1994improved} implemented in the Femtosoft 3.2.4 software (Femtosoft Technologies). The FROG measurements were further validated by comparing their frequency marginals with the autoconvolution of independently measured optical spectra (see SI Section S5). The laser source used in the FROG measurement was a titanium-sapphire laser (Coherent Chameleon Ultra II, repetition rate of 80 MHz) in combination with an OPO (APE, Chameleon Compact).

\subsection*{Author Contributions}
A.F., K.T., and I.S. conceived the idea. K.T. designed and fabricated the samples. A.F. performed the experiments, analyzed the data, and prepared the manuscript. R.S. and D.N. contributed to the methodology for pulse characterization. D.N. and I.S. supervised the project, and T.P., D.N., and I.S. provided funding and resources for the project. All authors discussed the results and contributed to the writing of the manuscript.

\subsection*{Acknowledgments}
The authors acknowledge financial support by the Deutsche Forschungsgemeinschaft (DFG, German Research Foundation) through the International Research Training Group IRTG 2675 “Meta-Active” (Project No. 437527638). This work was supported by the Free State of Thuringia and the European Social Fund Plus.

\subsection*{Conflicts of Interest}
The authors declare no conflicts of interest.

\subsection*{Data Availability Statement}
The data that support the findings of this study are available from the corresponding author upon reasonable request.
\medskip

\bibliographystyle{MSP}
\bibliography{advance}

@book{brabec2008strong,
  title={Strong field laser physics},
  author={Brabec, Thomas and Kapteyn, Henry},
  volume={1},
  year={2008},
  publisher={Springer}
}

@article{zewail2000femtochemistry,
  title={Femtochemistry: atomic-scale dynamics of the chemical bond using ultrafast lasers (Nobel Lecture)},
  author={Zewail, Ahmed H},
  journal={Angewandte Chemie International Edition},
  volume={39},
  number={15},
  pages={2586--2631},
  year={2000},
  publisher={Wiley Online Library}
}

@article{dudovich2002single,
  title={Single-pulse coherently controlled nonlinear Raman spectroscopy and microscopy},
  author={Dudovich, Nirit and Oron, Dan and Silberberg, Yaron},
  journal={Nature},
  volume={418},
  number={6897},
  pages={512--514},
  year={2002},
  publisher={Nature Publishing Group UK London}
}

@article{drexler2004ultrahigh,
  title={Ultrahigh-resolution optical coherence tomography},
  author={Drexler, Wolfgang},
  journal={Journal of biomedical optics},
  volume={9},
  number={1},
  pages={47--74},
  year={2004},
  publisher={Society of Photo-Optical Instrumentation Engineers}
}

@article{sidick1995ultrashort,
  title={Ultrashort-pulse second-harmonic generation. I. Transform-limited fundamental pulses},
  author={Sidick, Erkin and Knoesen, Andr{\'e} and Dienes, Andrew},
  journal={Journal of the Optical Society of America B},
  volume={12},
  number={9},
  pages={1704--1712},
  year={1995},
  publisher={Optical Society of America}
}

@article{lozovoy2005systematic,
  title={Systematic control of nonlinear optical processes using optimally shaped femtosecond pulses},
  author={Lozovoy, Vadim V and Dantus, Marcos},
  journal={ChemPhysChem},
  volume={6},
  number={10},
  pages={1970--2000},
  year={2005},
  publisher={Wiley Online Library}
}

@article{simon1989ultrashort,
  title={Ultrashort light pulses},
  author={Simon, John D},
  journal={Review of scientific instruments},
  volume={60},
  number={12},
  pages={3597--3624},
  year={1989},
  publisher={American Institute of Physics}
}

@article{wise2008high,
  title={High-energy femtosecond fiber lasers based on pulse propagation at normal dispersion},
  author={Wise, Frank W and Chong, Andy and Renninger, William H},
  journal={Laser \& Photonics Reviews},
  volume={2},
  number={1-2},
  pages={58--73},
  year={2008},
  publisher={Wiley Online Library}
}

@article{walmsley2001role,
  title={The role of dispersion in ultrafast optics},
  author={Walmsley, Ian and Waxer, Leon and Dorrer, Christophe},
  journal={Review of Scientific Instruments},
  volume={72},
  number={1},
  pages={1--29},
  year={2001},
  publisher={American Institute of Physics}
}

@article{dar2017chromatic,
  title={Chromatic dispersion compensation techniques and characterization of fiber Bragg grating for dispersion compensation},
  author={Dar, Aasif Bashir and Jha, Rakesh Kumar},
  journal={Optical and quantum electronics},
  volume={49},
  number={3},
  pages={108},
  year={2017},
  publisher={Springer}
}

@article{strickland1985compression,
  title={Compression of amplified chirped optical pulses},
  author={Strickland, Donna and Mourou, Gerard},
  journal={Optics communications},
  volume={55},
  number={6},
  pages={447--449},
  year={1985},
  publisher={Elsevier}
}

@article{brabec2000intense,
  title={Intense few-cycle laser fields: Frontiers of nonlinear optics},
  author={Brabec, Thomas and Krausz, Ferenc},
  journal={Reviews of Modern Physics},
  volume={72},
  number={2},
  pages={545},
  year={2000},
  publisher={APS}
}

@article{bor1985group,
  title={Group velocity dispersion in prisms and its application to pulse compression and travelling-wave excitation},
  author={Bor, Zs and Racz, B},
  journal={Optics communications},
  volume={54},
  number={3},
  pages={165--170},
  year={1985},
  publisher={Elsevier}
}

@article{treacy1969optical,
  title={Optical pulse compression with diffraction gratings},
  author={Treacy, Edmond},
  journal={IEEE Journal of quantum Electronics},
  volume={5},
  number={9},
  pages={454--458},
  year={1969},
  publisher={IEEE}
}

@article{dombi2005pulse,
  title={Pulse compression with time-domain optimized chirped mirrors},
  author={Dombi, Peter and Yakovlev, Vladislav S and O’Keeffe, Kevin and Fuji, Takao and Lezius, Matthias and Tempea, Gabriel},
  journal={Optics express},
  volume={13},
  number={26},
  pages={10888--10894},
  year={2005},
  publisher={Optical Society of America}
}

@article{chia2025multi,
  title={Multi-band chirped mirrors for enhanced dispersion management},
  author={Chia, Shih-Hsuan and Li, Yan-Cheng and Sun, Chi-Kuang and K{\"a}rtner, Franz X},
  journal={Optics \& Laser Technology},
  volume={180},
  pages={111525},
  year={2025},
  publisher={Elsevier}
}

@article{poole1994optical,
  title={Optical fiber-based dispersion compensation using higher order modes near cutoff},
  author={Poole, Craig D and Wiesenfeld, Jay M and DiGiovanni, DJ and Vengsarkar, Ashish M},
  journal={Journal of Lightwave Technology},
  volume={12},
  number={10},
  pages={1746--1758},
  year={1994},
  publisher={IEEE}
}

@article{gruner2005dispersion,
  title={Dispersion-compensating fibers},
  author={Gr{\"u}ner-Nielsen, Lars and Wandel, Marie and Kristensen, Poul and Jorgensen, Carsten and Jorgensen, Lene Vilbrad and Edvold, Bent and P{\'a}lsd{\'o}ttir, Bera and Jakobsen, Dan},
  journal={Journal of Lightwave Technology},
  volume={23},
  number={11},
  pages={3566},
  year={2005},
  publisher={OSA}
}

@article{weiner2000femtosecond,
  title={Femtosecond pulse shaping using spatial light modulators},
  author={Weiner, Andrew M},
  journal={Review of scientific instruments},
  volume={71},
  number={5},
  pages={1929--1960},
  year={2000},
  publisher={American Institute of Physics}
}

@article{lee2006optical,
  title={Optical Dispersion Compensator With $≫ $4000-ps/nm Tuning Range Using a Virtually Imaged Phased Array (VIPA) and Spatial Light Modulator (SLM)},
  author={Lee, G-H and Xiao, Shijun and Weiner, Andrew M},
  journal={IEEE photonics technology letters},
  volume={18},
  number={17},
  pages={1819--1821},
  year={2006},
  publisher={IEEE}
}

@article{xie2005dispersion,
  title={Dispersion compensation in high-speed optical coherence tomography by acousto-optic modulation},
  author={Xie, Tuqiang and Wang, Zhenguo and Pan, Yingtian},
  journal={Applied optics},
  volume={44},
  number={20},
  pages={4272--4280},
  year={2005},
  publisher={Optical Society of America}
}

@article{catanese2021acousto,
  title={Acousto-optic modulator pulse-shaper compression of octave-spanning pulses from a stretched hollow-core fiber},
  author={Catanese, Anthony and Kaufman, Brian and Cheng, Chuan and Jones, Eric and Cohen, Martin G and Weinacht, Thomas},
  journal={OSA Continuum},
  volume={4},
  number={12},
  pages={3176--3183},
  year={2021},
  publisher={Optical Society of America}
}

@article{gokce2014femtosecond,
  title={Femtosecond pulse shaping using the geometric phase},
  author={G{\"o}kce, Bilal and Li, Yanming and Escuti, Michael J and Gundogdu, Kenan},
  journal={Optics Letters},
  volume={39},
  number={6},
  pages={1521--1524},
  year={2014},
  publisher={Optical Society of America}
}

@article{rahimi2015femtosecond,
  title={Femtosecond pulse shaping by ultrathin plasmonic metasurfaces},
  author={Rahimi, Eesa and {\c{S}}endur, K{\"u}r{\c{s}}at},
  journal={Journal of the Optical Society of America B},
  volume={33},
  number={2},
  pages={A1--A7},
  year={2015},
  publisher={Optical Society of America}
}

@article{ossiander2021slow,
  title={Slow light nanocoatings for ultrashort pulse compression},
  author={Ossiander, Marcus and Huang, Y-W and Chen, Wei Ting and Wang, Zhenhao and Yin, Xinghui and Ibrahim, Yousef Ahmed and Schultze, Martin and Capasso, Federico},
  journal={Nature Communications},
  volume={12},
  number={1},
  pages={6518},
  year={2021},
  publisher={Nature Publishing Group UK London}
}

@article{geromel2023compact,
  title={Compact metasurface-based optical pulse-shaping device},
  author={Geromel, Ren{\'e} and Georgi, Philip and Protte, Maximilian and Lei, Shiwei and Bartley, Tim and Huang, Lingling and Zentgraf, Thomas},
  journal={Nano Letters},
  volume={23},
  number={8},
  pages={3196},
  year={2023}
}

@article{divitt2019ultrafast,
  title={Ultrafast optical pulse shaping using dielectric metasurfaces},
  author={Divitt, Shawn and Zhu, Wenqi and Zhang, Cheng and Lezec, Henri J and Agrawal, Amit},
  journal={Science},
  volume={364},
  number={6443},
  pages={890--894},
  year={2019},
  publisher={American Association for the Advancement of Science}
}

@article{chen2022synthesizing,
  title={Synthesizing ultrafast optical pulses with arbitrary spatiotemporal control},
  author={Chen, Lu and Zhu, Wenqi and Huo, Pengcheng and Song, Junyeob and Lezec, Henri J and Xu, Ting and Agrawal, Amit},
  journal={Science Advances},
  volume={8},
  number={43},
  pages={eabq8314},
  year={2022},
  publisher={American Association for the Advancement of Science}
}

@article{chen2023shaping,
  title={Shaping polarization within an ultrafast laser pulse using dielectric metasurfaces},
  author={Chen, Lu and Huo, P and Song, Junyeob and Wang, Zi and Xu, Ting and Zhu, Wenqi and Agrawal, Amit},
  journal={Optica},
  volume={10},
  number={1},
  pages={26--32},
  year={2023},
  publisher={Optica Publishing Group}
}

@article{tanaka2025femtosecond,
  title={Femtosecond pulse shaping with semiconductor Huygens' metasurfaces},
  author={Tanaka, Katsuya and Arslan, Dennis and Weissflog, Maximilian and Geib, Nils and Gerold, Kristin and Szeghalmi, Adriana and Ziegler, Mario and Eilenberger, Falk and Pertsch, Thomas and Schiek, Roland and others},
  journal={Advanced Optical Materials},
  volume={13},
  number={16},
  pages={2500135},
  year={2025},
  publisher={Wiley Online Library}
}

@article{decker2015high,
  title={High-efficiency dielectric Huygens’ surfaces},
  author={Decker, Manuel and Staude, Isabelle and Falkner, Matthias and Dominguez, Jason and Neshev, Dragomir N and Brener, Igal and Pertsch, Thomas and Kivshar, Yuri S},
  journal={Advanced Optical Materials},
  volume={3},
  number={6},
  pages={813--820},
  year={2015},
  publisher={Wiley Online Library}
}

@article{chen2023dispersion,
  title={Dispersion-engineered metasurfaces reaching broadband 90\% relative diffraction efficiency},
  author={Chen, Wei Ting and Park, Joon-Suh and Marchioni, Justin and Millay, Sophia and Yousef, Kerolos MA and Capasso, Federico},
  journal={Nature Communications},
  volume={14},
  number={1},
  pages={2544},
  year={2023},
  publisher={Nature Publishing Group UK London}
}

@article{barati2022optical,
  title={Optical pulse compression assisted by high-Q time-modulated transmissive metasurfaces},
  author={Barati Sedeh, Hooman and Salary, Mohammad Mahdi and Mosallaei, Hossein},
  journal={Laser \& Photonics Reviews},
  volume={16},
  number={5},
  pages={2100449},
  year={2022},
  publisher={Wiley Online Library}
}

@book{malacara2007optical,
  title={Optical shop testing},
  author={Malacara, Daniel and others},
  volume={3},
  year={2007},
  publisher={Wiley-Interscience Hoboken, NJ}
}

@book{diels2006ultrashort,
  title={Ultrashort laser pulse phenomena},
  author={Diels, Jean-Claude and Rudolph, Wolfgang and Rudolph, Wolfgang},
  volume={2},
  year={2006},
  publisher={Elsevier}
}

@article{o2001highly,
  title={Highly simplified device for ultrashort-pulse measurement},
  author={O’shea, Patrick and Kimmel, Mark and Gu, Xun and Trebino, Rick},
  journal={Optics letters},
  volume={26},
  number={12},
  pages={932--934},
  year={2001},
  publisher={Optical Society of America}
}

@article{delong1994pulse,
  title={Pulse retrieval in frequency-resolved optical gating based on the method of generalized projections},
  author={DeLong, Kenneth W and Fittinghoff, David N and Trebino, Rick and Kohler, Bern and Wilson, Kent},
  journal={Optics letters},
  volume={19},
  number={24},
  pages={2152--2154},
  year={1994},
  publisher={Optical Society of America}
}

@article{delong1994improved,
  title={Improved ultrashort pulse-retrieval algorithm for frequency-resolved optical gating},
  author={DeLong, Kenneth W and Trebino, Rick},
  journal={Journal of the Optical Society of America A},
  volume={11},
  number={9},
  pages={2429--2437},
  year={1994},
  publisher={Optical Society of America}
}

\clearpage


\section*{Supplementary material}

# Supplementary Information

## In-line Dispersion Control of Femtosecond Laser Pulses by Multilayer Huygens' Metasurfaces

*Anna Fitriana,* Katsuya Tanaka, Roland Schiek, Thomas Pertsch, Dragomir Neshev, Isabelle Staude*

## S1 Metasurface Geometry

Each Huygens' metasurface consists of silicon (Si) nanodisks with radius $r$ and height $h$ arranged on a square lattice with period $p$. The Si nanodisks were fabricated on top of a glass substrate and embedded in a glassy matrix via spin coating (ICI-200). The effective geometrical parameters of the fabricated metasurfaces ($r = 251$ nm, $h = 225$ nm, and $p = 754$ nm) were obtained by numerically fitting the measured complex transmission response in COMSOL Multiphysics 6.2.

To this end, we implemented a numerical optimization scheme to quantitatively retrieve the geometrical parameters of the nanodisks by minimizing the deviation between the numerically simulated and the experimentally measured complex transmission spectra. The optimization problem was formulated as a nonlinear least squares problem (MATLAB, Optimoptions package) using a Levenberg-Marquadt algorithm [1].

For comparison, Figure S1 shows the scanning electron micrographs (SEMs) of a fabricated Huygens' metasurface, where the reported radius and height are 285 and 243 nm, respectively. The discrepancy between the optimized and measured geometries may arise due to an oxide layer forming around the nanodisks during fabrication [2], which was not taken into account in the numerical optimization.

Figure S2a shows the absolute transmittance $T$, calculated as the ratio of the transmitted intensities through the $N$-stack metasurfaces to the incident intensity, $I_{\text{trans}}/I_{\text{in}}$, while the relative

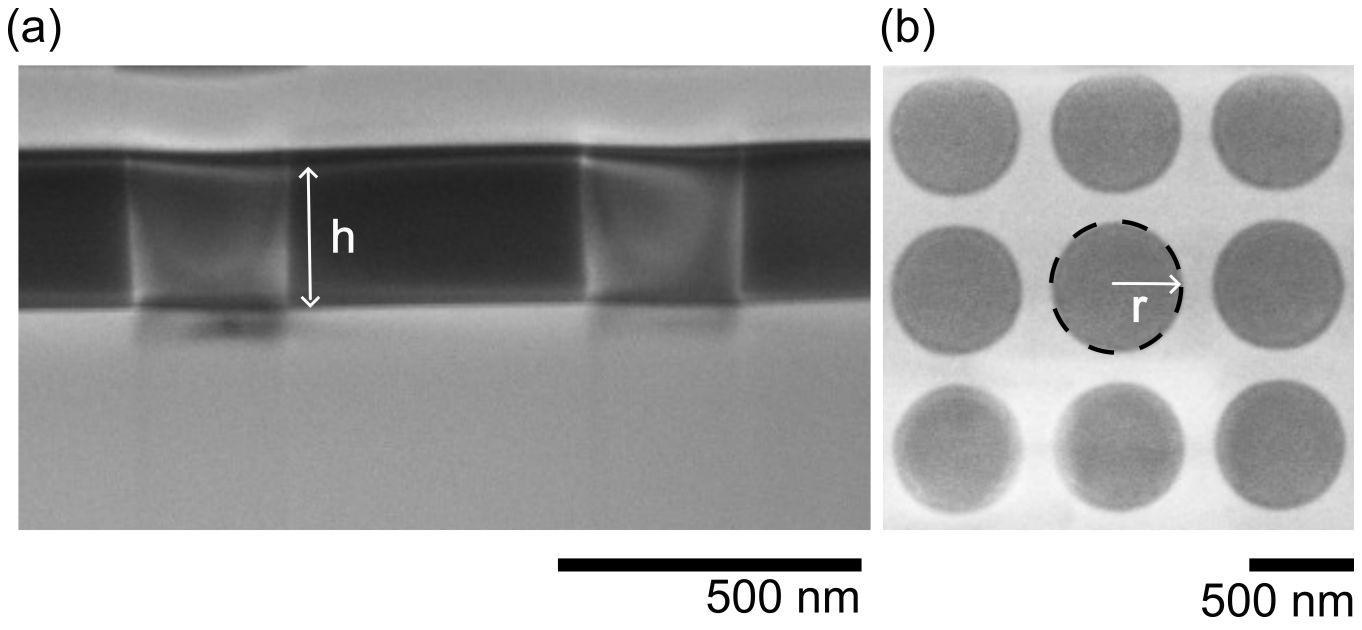


Figure S1: (a) Side and (b) top views of scanning electron micrograph images of the Huygens' silicon nanostructures, obtained by a focused ion beam (FIB) cross sectioning. The radius and height of the silicon nanodisk are denoted by $r$ and $h$, respectively.

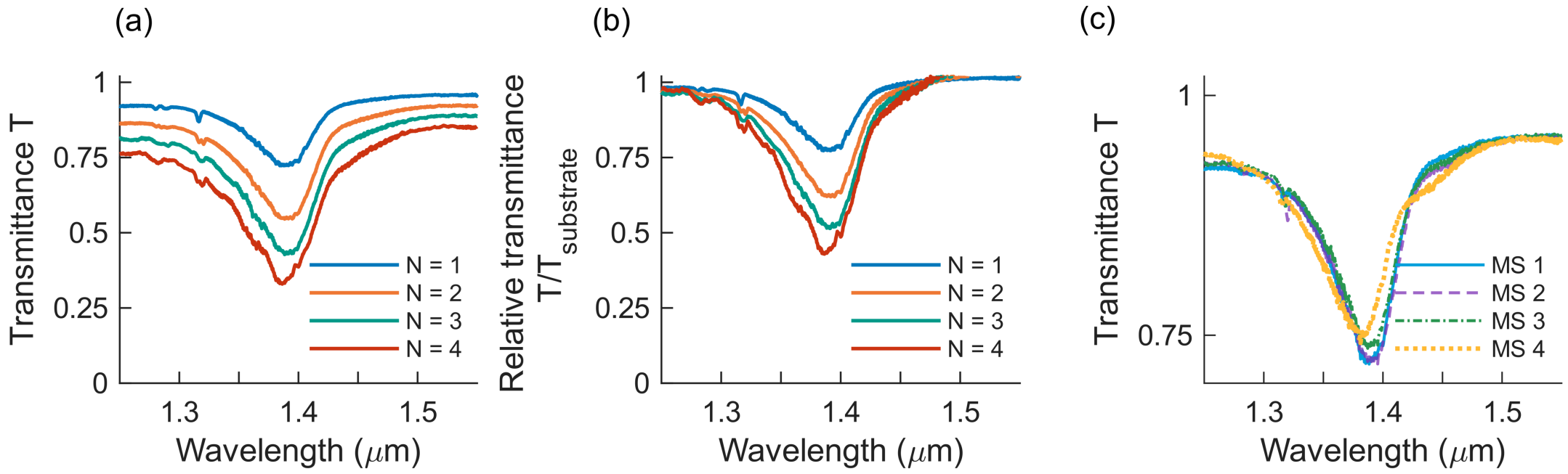


Figure S2: Measured transmittance of $N$-stack of Huygens' metasurfaces (a) with the input intensity used as the reference; (b) relative to the transmittance of $N$-stack of glass substrates; and (c) transmittance of the individual metasurface (MS).

transmittance $T_r$ in Figure S2b was obtained by normalizing $T$ to the transmittance of the corresponding substrate stack $T_{\text{substrate}}$. Both measurements show a systematical reduction in transmittance values with increasing number of layers. The transmittance of the individual metasurface given in Figure S2c exhibits a nearly identical spectral response for each metasurface.

## S2 Fabrication

The silicon metasurfaces were fabricated using a standard top-down nanofabrication process based on electron-beam lithography and reactive ion etching. First, commercially available amorphous silicon thin films on glass substrates (Tafelmaier Dünnschicht Technik GmbH) were etched to the required thickness using argon ion-beam etching (Oxford Instruments Plasma Technology, Ionfab 300). Afterward, a conductive chromium layer was deposited by ion-beam sputtering using the same system. Next, a 100 nm-thick negative electron-beam resist (EN038, Tokyo Ohka Kogyo Co., Ltd.) was spin-coated onto the sample surface. The resist was exposed using a variable-shaped electron-beam lithography system (Vistec SB 350OS) to define the desired nanopatterns,

followed by development in OPD 4262 for 30 seconds. The resist pattern was then transferred into the chromium layer by ion-beam etching (Oxford Instruments Plasma Technology, Ionfab 300LC), forming a hard mask for the subsequent silicon etching step. The silicon layer was structured using reactive ion etching with an inductively coupled plasma system (RIE-ICP, Sen-tech SI-500 C). Finally, the remaining resist and chromium mask were removed. The resist was dissolved in acetone, while the chromium layer was removed using a ceric ammonium nitrate-based solution in a megasonic bath at 50 °C. The fabricated nanostructures were subsequently embedded in spin-on-glass (ICI-200) by spin coating, followed by thermal annealing at 200 °C.

## S3 Resonant Phase in Huygens' Regime

The optical response of individual nanodisk can be represented by a pair of electric and magnetic dipoles that follow a Lorentzian profile where the resonant electric and magnetic dipoles located at $\omega_{\mathrm{e}}$ and $\omega_{\mathrm{m}}$ and the corresponding damping parameters are given by $\gamma_{\mathrm{e}}$ and $\gamma_{\mathrm{m}}$. In a loss-less Huygens' system, the field transmission and reflection were analytically derived based on a coupled discrete dipole approach [3] given as follows,

$$t(\omega) = 1 + \frac{2\mathrm{i}\gamma_{\mathrm{e}}\omega}{\omega_{\mathrm{e}}^2 - \omega^2 - 2\mathrm{i}\gamma_{\mathrm{e}}\omega} + \frac{2\mathrm{i}\gamma_{\mathrm{m}}\omega}{\omega_{\mathrm{m}}^2 - \omega^2 - 2\mathrm{i}\gamma_{\mathrm{m}}\omega} \tag{S1}$$

and

$$r(\omega) = \frac{2\mathrm{i}\gamma_{\mathrm{e}}\omega}{\omega_{\mathrm{e}}^2 - \omega^2 - 2\mathrm{i}\gamma_{\mathrm{e}}\omega} - \frac{2\mathrm{i}\gamma_{\mathrm{m}}\omega}{\omega_{\mathrm{m}}^2 - \omega^2 - 2\mathrm{i}\gamma_{\mathrm{m}}\omega} \tag{S2}$$

where $t(\omega)$ and $r(\omega)$ are frequency dependent transmission and reflection coefficients, respectively. Considering the spectral overlap between electric and magnetic resonances in Huygens' regime, we set $\omega_{\mathrm{e}} = \omega_{\mathrm{m}} \equiv \omega_0$. In practice, however, the Huygens' condition is not perfectly satisfied, and a small residual reflectance (Equation (S2)) may occur due to unequal damping rates of the electric and magnetic dipoles, i.e., $\gamma_{\mathrm{e}} \neq \gamma_{\mathrm{m}}$. Further, in a regime very close to the resonance, $\omega \approx \omega_0$, the denominator term becomes $\omega_0^2 - \omega^2 = 2\omega_0(\omega_0 - \omega)$. Also, by taking $(\omega_0 - \omega)^2 \ll \gamma_{\mathrm{e}}^2$ or $\gamma_{\mathrm{m}}^2$ the complex transmission in Equation (S1) can be simplified as,

$$t(\omega) = 1 + \frac{4\mathrm{i}\gamma_\mathrm{e}\omega\omega_0(\omega_0-\omega) - 4\gamma_\mathrm{e}^2\omega^2}{4\gamma_\mathrm{e}^2\omega^2} + \frac{4\mathrm{i}\gamma_\mathrm{m}\omega\omega_0(\omega_0-\omega) - 4\gamma_\mathrm{m}^2\omega^2}{4\gamma_\mathrm{m}^2\omega^2} \tag{S3}$$

$$= -1 + \mathrm{i}\left[\frac{\omega_0-\omega}{\gamma_\mathrm{e}} + \frac{\omega_0-\omega}{\gamma_\mathrm{m}}\right] \tag{S4}$$

From Equation (S3), we can formulate the corresponding transmission phase of the Huygens' metasurface $\varphi(\omega)$ by introducing an effective damping parameter $\gamma$, defined by

$$\varphi(\omega) = \arctan\left[\frac{\omega-\omega_0}{\gamma/2}\right], \gamma = \frac{2\gamma_\mathrm{e}\gamma_\mathrm{m}}{\gamma_\mathrm{e}+\gamma_\mathrm{m}} \tag{S5}$$

We first fitted the measured phase from the interferometry measurement with the analytical function in Equation (S5). Then, we calculated the group delay dispersion through $\frac{\mathrm{d}^2\varphi(\omega)}{\mathrm{d}\omega^2}$, where $\varphi(\omega)$ here represents the fitted spectral phase.

## S4 Spectral Phase Interferometry

The spectral phase interferometry records the interferogram resulting from the superposition of a beam transmitted through the sample arm and a beam whose phase is retarded by a geometric phase shifter [4]. The corresponding beam paths are referred to as reference and sample arms, respectively. The phase shifter consists of a half wave plate (HWP) mounted on a motorized rotation stage, which is placed between two fixed quarter wave plates (QWPs) with the fast axis of both QWPs oriented at $\frac{\pi}{4}$ relative to the x-axis (see Figure 2 in the main text).

The change in the complex electric field vector transmitted by a retarder can be generally expressed in the following standard Jones matrix, e.g., HWP $= \begin{bmatrix} 1 & 0 \\ 0 & -1 \end{bmatrix}$ and QWP $= \begin{bmatrix} 1 & 0 \\ 0 & -\mathrm{i} \end{bmatrix}$ for a half wave plate and quarter wave plate, respectively. When the fast axis of the HWP or QWP is rotated by an angle $\theta$ relative to the x-axis, the final Jones matrix is transformed using a coordinate rotation matrix $\mathrm{R}(\theta)$, i.e.,

$$J_{\mathrm{HWP}}(\theta) = R(-\theta)\begin{bmatrix}1 & 0\\ 0 & -1\end{bmatrix}R(\theta) \tag{S6}$$

$$J_{\mathrm{QWP}}(\theta) = R(-\theta)\begin{bmatrix}1 & 0\\ 0 & -\mathrm{i}\end{bmatrix}R(\theta), \tag{S7}$$

with R($\theta$) expressed as

$$\mathrm{R}(\theta) = \begin{bmatrix}\cos(\theta) & \sin(-\theta)\\ \sin(\theta) & \cos(\theta)\end{bmatrix}. \tag{S8}$$

Taking the fast axis of the two QWPs at an angle of $\theta = +\frac{\pi}{4}$ and the HWP at an angle of $\Delta$, the net phase retardation in the reference arm can be written as follows

$$J(\Delta) = J_{\mathrm{QWP}}\Big(\frac{\pi}{4}\Big)J_{\mathrm{HWP}}(\Delta)J_{\mathrm{QWP}}\Big(\frac{\pi}{4}\Big) = \begin{bmatrix}\mathrm{i}e^{-2\mathrm{i}\Delta} & 0\\ 0 & -\mathrm{i}e^{2\mathrm{i}\Delta}\end{bmatrix}. \tag{S9}$$

The complex transmission matrix of the sample is defined by $t_s = \begin{bmatrix}t_{xx}(x,y) & t_{yx}(x,y)\\ t_{xy}(x,y) & t_{yy}(x,y)\end{bmatrix}$. Assume that, after the first beam splitter, the transmittances of the beams propagating through the sample and reference arms are $T_{\mathrm{S}}$ and $T_{\mathrm{R}}$, respectively, where $T_{\mathrm{S}} = 1 - T_{\mathrm{R}}$. Within the paraxial approximation, the beam impinging on the detector can be treated as a quasi plane wave where the wave propagating along the z-direction has a fast-oscillating carrier wave $e^{-\mathrm{i}kz}$. Consequently, the interference of electric fields on the detector $\boldsymbol{E}$ is further formulated by

$$\boldsymbol{E}(x,y,\Delta) = \Big(\sqrt{1-T_R}\begin{bmatrix}t_{xx}(x,y) & t_{yx}(x,y)\\ t_{xy}(x,y) & t_{yy}(x,y)\end{bmatrix} + \sqrt{T_R}\begin{bmatrix}\mathrm{i}e^{-2\mathrm{i}\Delta} & 0\\ 0 & -\mathrm{i}e^{2\mathrm{i}\Delta}\end{bmatrix}\Big)\boldsymbol{E}_0(x,y)e^{-\mathrm{i}kz} \tag{S10}$$

where $\mathbf{E}_0$ is the electric field of the laser source. If we only consider x-linear polarized input, i.e., $\boldsymbol{E}_0 = \begin{bmatrix}1\\ 0\end{bmatrix}E_0$, the corresponding intensity can be written as

$$I(x,y,\Delta) = \frac{1}{2c_0\mu_0}|\boldsymbol{E}(x,y,\Delta)|^2 \tag{S11}$$

$$= \frac{|E_0|^2}{2c_0\mu_0}\Big(T_R + (1-T_R)(|t_{xx}|^2 + |t_{xy}|^2) + 2\sqrt{T_R(1-T_R)}|t_{xx}|\sin(2\Delta + \arg(t_{xx})\Big). \tag{S12}$$

By evaluating Equation (S12), the complex transmission coefficients of the sample, $t_{xx}$, can be resolved from a series of interferograms acquired at different phase delay $\Delta$. In principle, at least three phase steps of $\Delta$ are required to uniquely determine the complex transmission coefficients [5, 6, 7]. In our scheme, we employed a phase shifting interferometry with 16 phase steps, where the interferograms are recorded at phase delay $\Delta_j = \frac{2\pi}{16} j$ with $j = \{0, 1, 2, .....15\}$. In general, having more phase steps improves the retrieval accuracy by suppressing the influence of noises on the resolved phase, which can be caused by experimental fluctuations, e.g., mechanical and turbulence drift, detector noises, and laser intensity variations [8].

## S5 Marginal Analysis

Frequency-resolved optical gating (FROG) marginals are critical tools to examine the consistency between different methods of pulse characterization, where the information obtained from the FROG results must be consistent with the independently measured spectrum and autocorrelation. As such, a marginal analysis allows for detecting possible sources of error in the measurement process [9]. Moreover, it helps to ensure that the iterative retrieval algorithm is converging into a physically valid solution.

In our study, we calculated the FROG frequency marginals and compared them with independently measured spectra. Firstly, the frequency marginal $M_\omega$ was defined by an integration of the FROG spectrogram $I_{\mathrm{FROG}}(\omega, \tau)$ across delay time $\tau$, as given in the following expression,

$$M_\omega(\omega) = \int d\tau I_{\mathrm{FROG}}(\omega, \tau). \tag{S13}$$

$M_\omega$ should agree with the autoconvolution of the measured spectrum $S$,

$$S * S(\omega) = \int d\omega' S(\omega') S(\omega - \omega'). \tag{S14}$$

Figure S3 shows a comparison between frequency marginals evaluated from the measured FROG trace and the calculated autoconvolution of the measured spectra. The excellent overall agree-

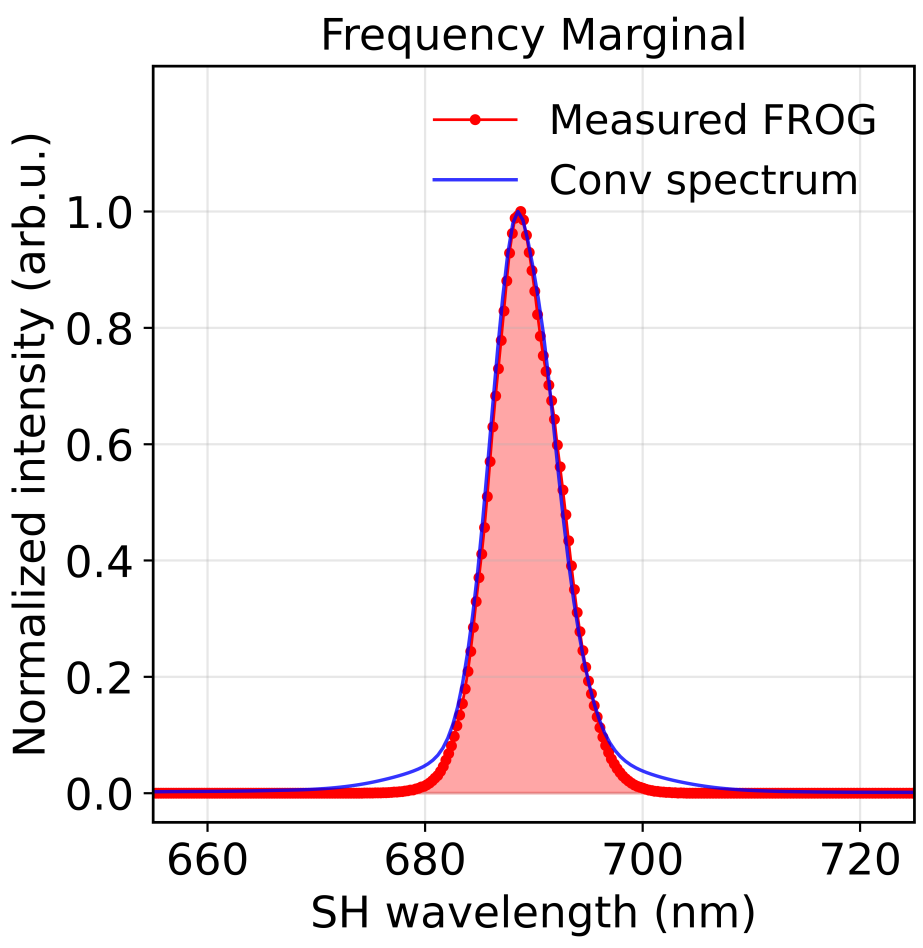


Figure S3: Marginal analysis of the FROG spectrograms for an input pulse centered at 1.375 ± 0.003 µm. Comparison between the measured FROG frequency marginal (red) and the autoconvolution of the independently measured spectrum (blue). The horizontal axis is the second harmonic (SH) wavelength.

ment underpins the reliability of our FROG measurements.

## S6 Intensity Autocorrelation

To characterize the temporal pulse duration of femtosecond laser pulses, we first performed interferometric (fringe-resolved) autocorrelation measurements at different signal central wavelengths (see Figure S4a). The corresponding intensity autocorrelation traces are obtained by filtering out the rapidly oscillating interference fringes from the interferometric autocorrelation signal. The filtering was implemented numerically in the Fourier domain (see Figure S4c). Notably, the extracted intensity autocorrelation (see Figure S4b) still includes the constant background contribution where the normalized signal approaches 1 in the limit of large delay $\tau \rightarrow \infty$.

The pulse duration is then retrieved by independently calculating the intensity autocorrelation signal from some pulse profiles to match the measured trace. Figure S4b shows that the autocorrelation of a Gaussian envelope provides an excellent agreement. Therefore, the actual pulse duration $\tau$ is evaluated at the full width at half maximum of that Gaussian pulse, which also specifically leads to the relation $\tau = 0.707\tau_{\mathrm{IAC}}$ where $\tau_{\mathrm{IAC}}$ is the width of the autocorrelation trace.

Additionally, while an ideally collinear interferometric autocorrelation exhibits an 8:1 peak-to-background ratio, however, our experiment obtains a ratio of 3.6:1 in Figure S4b. This is asso-

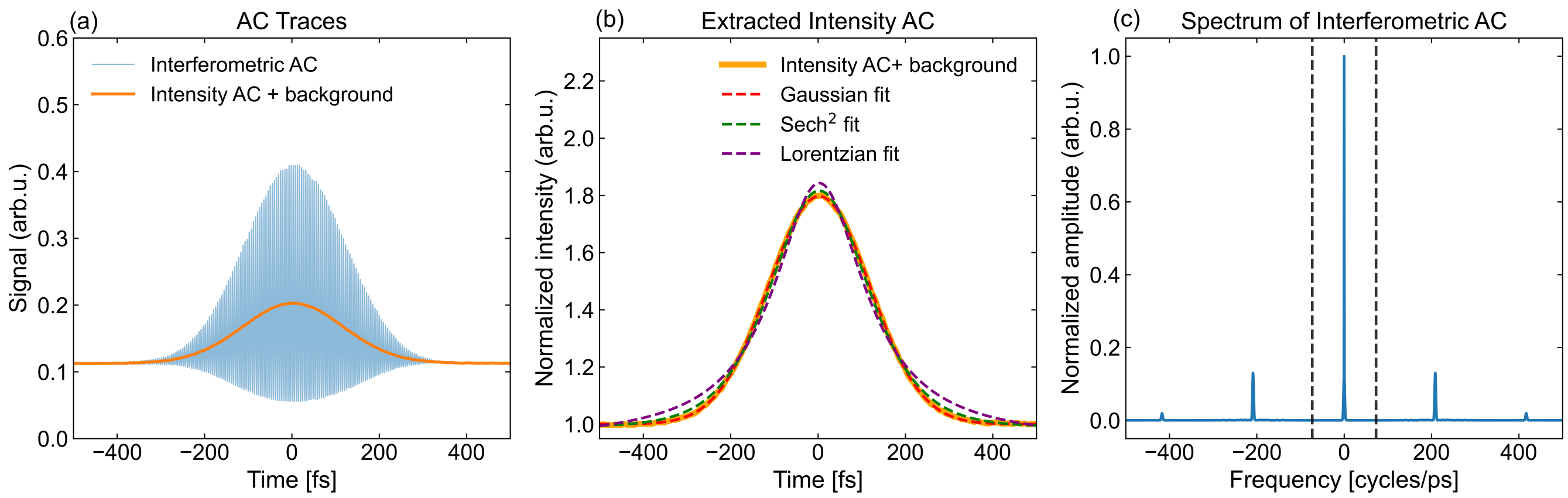


Figure S4: Autocorrelation measurement of an input pulse. (a) Fringe-resolved (interferometric) autocorrelation signals (blue) and the corresponding filtered trace, obtained by filtering out the fast oscillatory components (orange). (b) Comparisons with modeled autocorrelation using different pulse shape. A Gaussian pulse envelope provides the best agreement with the measured results. (c) Example of numerical filtering to extract the intensity autocorrelation is implemented in the Fourier domain.

ciated with an incomplete spatial overlap of the two pulses in the autocorrelator nonlinear crystal, which further leads to a less efficient nonlinear phase match [10]. Consequently, the intensity autocorrelation trace shows a 1.8:1 peak-to-background ratio, which should be 3:1 peak-to-background ratio ideally in a collinear autocorrelation.

## S7 FROG Error

The FROG error ($G$) is calculated from the root-mean-square (RMS) difference between the experimentally measured ($I_{\text{meas}}$) and retrieved traces ($I_{\text{ret}}$), written as [11]

$$G = \sqrt{\frac{1}{N_F^2}\Sigma_i\Sigma_j\left|I_{\text{meas}}(\omega_i,\tau_j) - \mu I_{\text{ret}}(\omega_i,\tau_j)\right|^2}. \tag{S15}$$

$N_{\text{F}}$ is the grid number in the FROG retrieval and $\mu$ is the scaling factor defined as $\mu = \frac{\Sigma_{i,j} I_{\text{meas}_{i,j}} I_{\text{ret}_{i,j}}}{\Sigma_{i,j} I^2_{\text{ret}_{i,j}}}$. In our FROG retrieval process, we employed a 512×512 grid and set a delay resolution of 12 fs per grid point and a spectral resolution of 0.183 THz per grid point

## References

[1] J. J. Moré, In *Numerical analysis: proceedings of the biennial Conference held at Dundee, June 28–July 1, 1977*. Springer, **2006** 105–116.

[2] T. Bucher, A. Vaskin, R. Mupparapu, F. J. Löchner, A. George, K. E. Chong, S. Fasold, C. Neumann, D.-Y. Choi, F. Eilenberger, et al., *ACS photonics* **2019**, *6*, 4 1002.

[3] M. Decker, I. Staude, M. Falkner, J. Dominguez, D. N. Neshev, I. Brener, T. Pertsch, Y. S. Kivshar, *Advanced Optical Materials* **2015**, *3*, 6 813.

[4] P. Hariharan, M. Roy **1992**.

[5] J. H. Bruning, D. R. Herriott, J. E. Gallagher, D. P. Rosenfeld, A. D. White, D. J. Brangaccio, *Applied optics* **1974**, *13*, 11 2693.

[6] J. Schwider, R. Burow, K.-E. Elssner, J. Grzanna, R. Spolaczyk, K. Merkel, *Applied optics* **1983**, *22*, 21 3421.

[7] P. Hariharan, B. F. Oreb, T. Eiju, *Applied optics* **1987**, *26*, 13 2504.

[8] D. Malacara, et al., *Optical shop testing*, volume 3, Wiley-Interscience Hoboken, NJ, **2007**.

[9] I. Amat-Roldán, I. G. Cormack, P. Loza-Alvarez, E. J. Gualda, D. Artigas, *Optics express* **2004**, *12*, 6 1169.

[10] R. Grunwald, U. Griebner, E. Nibbering, A. Kummrow, M. Rini, T. Elsaesser, V. Kebbel, H.-J. Hartmann, W. Jüptner, *Journal of the Optical Society of America A* **2001**, *18*, 11 2923.

[11] R. Trebino, K. W. DeLong, D. N. Fittinghoff, J. N. Sweetser, M. A. Krumbügel, B. A. Richman, D. J. Kane, *Review of Scientific Instruments* **1997**, *68*, 9 3277.



\end{document}